\documentclass{emulateapj}  

\usepackage{subfigure}

\newcommand{\galics}{GalICS }

\newcommand{\sachant}[2]{\left( #1 \left| #2 \right.\right)}
\newcommand{\norm}[1]{\left\| {#1} \right\|}
\newcommand{\paren}[1]{ {\left( {#1} \right)} }
\def \vr {{\bf r}}
\def \vv {{\bf v}}

\shorttitle{MoLUSC}
\shortauthors{Sousbie, Courtois, Bryan \& Devriendt}

\begin{document} 

\title{MoLUSC: A MOck Local Universe Survey Constructor}

\author{T. Sousbie\altaffilmark{1}, H. Courtois\altaffilmark{1, 2}, G. Bryan\altaffilmark{3}, J. Devriendt\altaffilmark{1}}

\altaffiltext{1}{Centre de Recherche Astrophysique de Lyon (CRAL), Universit\'e Lyon I, 9 avenue Charles Andr\'e, 69561 Saint Genis Laval Cedex, France}
\altaffiltext{2}{Institute for Astronomy, 2680 Woodlawn Drive, Honolulu, HI 96822}
\altaffiltext{3}{Department of Astronomy, Columbia Unviersity, Pupin Physics Laboratories, New York, NY 10027}


\begin{abstract}
This paper presents MoLUSC, a new method for generating mock galaxy
catalogs from a large scale ($\approx 1000^3$ Mpc$^3$) dark matter
simulation, that requires only modest CPU time and memory allocation.
The method uses a small-scale ($\approx 256^3$ Mpc$^3$) dark matter simulation
on which the \galics semi-analytic code has been run in order to define
the transformation from dark matter density to galaxy density transformation
using a probabilistic treatment.  MoLUSC is then applied
to a large-scale dark matter simulation in order to produce a realistic
distribution of galaxies and their associated spectra.
This permits the fast generation of large-scale mock surveys using
relatively low-resolution simulations.
We describe various tests which have been conducted to validate the method,
and demonstrate a first application to generate a mock Sloan Digital Sky
Survey redshift survey.
\end{abstract}
     



\section{Introduction}

Cosmological simulations are of great importance for studies of the formation of large
scale structure in the universe. They permit us to objectively test cosmological models
against observations, and in the past have helped to rule out models.  However, analysis
of these surveys is difficult due to the biases and limitations that are inherent in any
observational survey of the galaxy distribution in the universe.  In order to quantify those
biases and to make detailed comparisons to theoretical models, we need to generate
realistic mock galaxy surveys.  This is particularly difficult for very large surveys
such as the Sloan Digital Sky Survey (SDSS) which survey a substantial volume of the
universe.

A common way to generate mock catalogs is to carry out simulations which model a
volume which is at least as large as the survey in question, while at the same time 
resolving the smallest dark matter halos that can host galaxies of interest (e.g.,
Jing et al 1998; Yan et al. 2003).  Once halos are identified in such a simulation,
it can be populated using techniques such as the Halo Occupation Distribution 
(Peacock \& Smith 2000; Zhao et al 2002). However, for the largest surveys 
which can span thousands of Mpc, it is often prohibitively expensive to
run such simulations, both in terms of memory and cpu.  

Instead, we describe a method which does not identify individual halos, but
instead uses a smaller-scale, but higher-resolution simulation to constrain the relation
between dark-matter density and galaxy density.  The higher-resolution
simulation does resolve all the relevant dark matter halos and so it can use
a more accurate method to populate the halos.  In this paper, we use the
\galics semi-analytic model on this smaller-scale simulation, but in principal
any technique to associate halos with galaxies could be used.  Our method
is similar in spirit to Cole et al (1998) and Hamana et al (2002) but our
method of computing the effective bias is more sophisticated (as it is based
on a full-blown semi-analytic method).


The large scale ($\approx 1000^3$ Mpc$^3$) cosmological dark matter (DM) 
simulations presented in this paper
were run using the GADGET-2 public code \cite{Springel05}.
GADGET-2 is a massively parallel code for hydrodynamical cosmological simulations,
although we use only its dark-matter capabilities.

The MoLUSC treatment, i.e. converting DM particules into a distribution with galaxy properties
was made using the GALICS public mock galaxy catalogs. 
The GalICS project \cite{Hatton03} describes hierarchical galaxy formation with the 
so-called ``hybrid'' approach. It uses the outputs of large cosmological N-body simulations 
to get a more realistic description of dark matter halos, and a semi-analytic model to
describe the baryons. Because it keeps a record of the spatial and dynamical
information, the hybrid approach opens the way to a detailed treatment of galaxy
interaction and merging. The GalICS model explicitly intends to 
address the issue of the high-redshift star formation rate history in a multi-wavelength
prospect, from the ultraviolet to the sub-millimeter range.

\section{Generating galaxies from dark matter simulations}

In the following, we make the assumption that the galaxy spatial distribution is
mainly affected by the underlying dark matter distribution and that all other
processes influencing it can be considered as stochastic and have no
significant impact on the large scale clustering properties
Let $S_l$ be a large scale dark-matter only simulation,
$S_s$ a small scale high resolution dark-matter only simulation and $G_s$ the
resulting galaxy distribution obtained by using GalICS on $S_s$. The process
used by MoLUSC to generate
 a galaxy distribution $G^{\star}_l$ out of $S_l$ consists of two main steps:

\begin{enumerate}
\item The computation of the bias between dark matter and galaxy distribution in
  $S_s$ and $G_s$. This is achieved by:
  \begin{enumerate}
  \item Sampling $S_s$ and $G_s$ density fields over {\em identical} grids
    ($\rho_{S_s}\paren{\vr_i}$ and $\rho_{G_s}\paren{\vr_i}$ hereafter).
  \item Computing the probability
    $P\sachant{\rho_{G_s}\paren{\vr_i}}{\rho_{S_s}\paren{\vr_i}}$ that for a
    given grid node $i$, a dark matter density $\rho_{S_s}\paren{\vr_i}$ and a
    galaxy density $\rho_{G_s}\paren{\vr_i}$ would be measured.
  \item Computing the probability that, for a given value of
    $\rho_{G_s}\paren{\vr}$ and $\rho_{S_s}\paren{\vr}$, a given spectrum would
    be assigned to a galaxy located in $\vr$.
  \end{enumerate}
\item The generation of a galaxy distribution $G^{\star}_l$ that
  respects the probability distributions computed in the first step while
  following the large scale structure distribution of $S_l$. This result is
  obtained by:
  \begin{enumerate}
    \item Sampling the density field of $S_l$ over a grid
      ($\rho_{S_l}\paren{\vr_i}$ hereafter).
    \item Building a density field $\rho_{G_l^\star}\paren{\vr_i}$ from
      $\rho_{S_l}\paren{\vr_i}$ and the probability distribution
      $P\sachant{\rho_{G_s}\paren{\vr_i}}{\rho_{S_s}\paren{\vr_i}}$.
    \item Creating a discrete galaxy distribution whose sampled density field
      is $\rho_{G_l^\star}\paren{\vr_i}$\\
  \end{enumerate}
\end{enumerate}

\subsection{Bias analysis}

As explained above, the dark matter to galaxy distribution bias is computed
from the sampled density fields of  $S_s$ and $G_s$. There
exist many efficient ways for sampling a density field from a discrete point
distribution. In this case, we want to keep track
of which galaxy contributed to which sampling grid node in order to be able to
recover the spectral information. The chosen method uses a
truncated Gaussian kernel and consists of considering the $i^{th}$ particle as a density
cloud $W\paren{\vr-\vr_i}$ centered on the particle location $\vr_i$. For a cubic sampling grid
with cell size $\sigma h^{-1}$ Mpc, the number density at the $k^{th}$ node is then
given by:
\begin{equation}
n\paren{\vr_k}=\sum_{i=0}^N W\paren{\vr_k-\vr_i},
\label{eq_density_nb}
\end{equation} 
and its mass density is:
\begin{equation}
\rho\paren{\vr_k}=\sum_{i=0}^N m_i W\paren{\vr_k-\vr_i},
\label{eq_density_mass}
\end{equation}
where $m_i$ is the mass of the $i^{th}$ particle and $N$ the total number of
particles.\\

The kernel function is chosen to be a truncated Gaussian of the form:
 \begin{equation}
W\paren{\vr}=\frac{A}{\paren{4\pi L^2}^{3/2}}\exp\paren{-\frac{\norm{\vr}}{2L^2}}\,\Pi\paren{\Delta\sigma-\norm{\vr}},
\label{eq_gaussian}
\end{equation}
where $A$ is a normalization constant, $L$ is the smoothing length, $\Pi$ is
the Heavyside function and $\Delta$ sets the truncation length. Given the
infinite extend of the Gaussian function, the kernel is truncated in order to
reduce the computation time. Experiments show that choosing a sampling length
equal to the smoothing length $\sigma = L$ (all information being  wiped out at
scales smaller than $L$) and setting the truncation length
to $\Delta = 5$ provides good results, the error on the measurement compared to
an infinite extend kernel being of order $10^{-7}$ for a homogeneous field.\\

\begin{figure}[htp]
  \centering
  \subfigure[$z=3$]{\includegraphics[width=6.5cm]{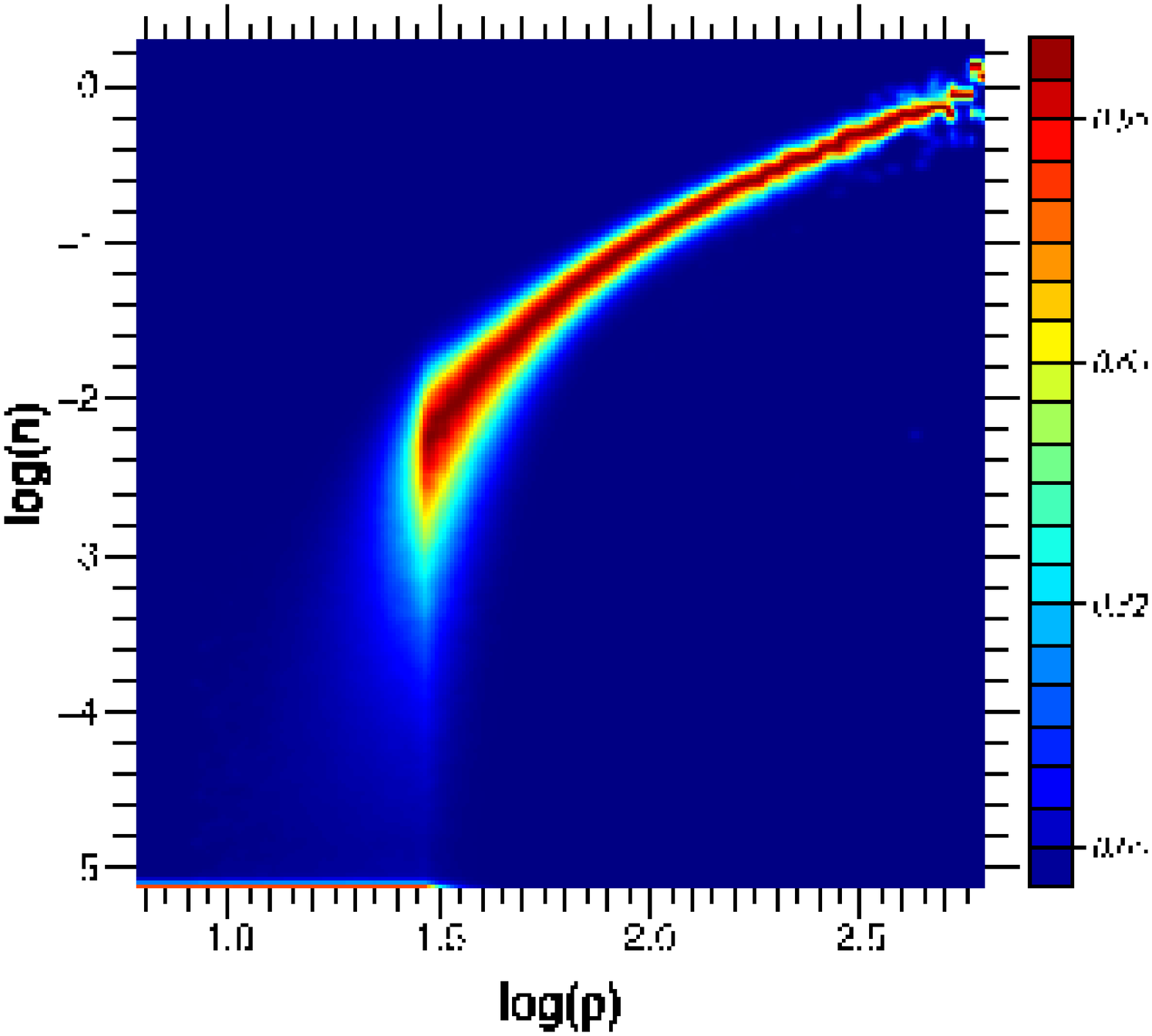}\label{fig_molusc_map_z3}}\\
  \centering
  \subfigure[$z=1$]{\includegraphics[width=6.5cm]{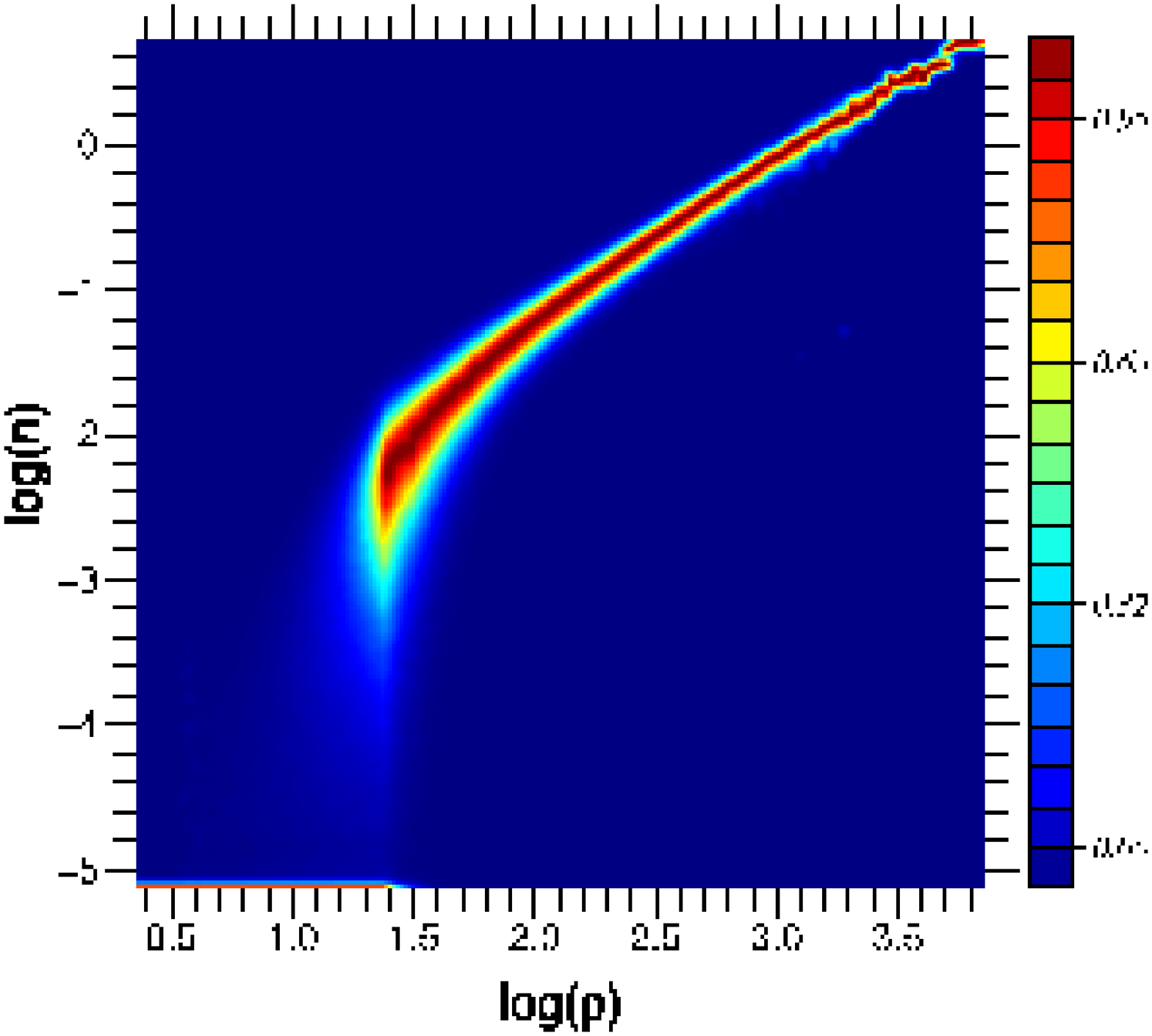}\label{fig_molusc_map_z1}}\\
  \centering
  \subfigure[$z=0$]{\includegraphics[width=6.5cm]{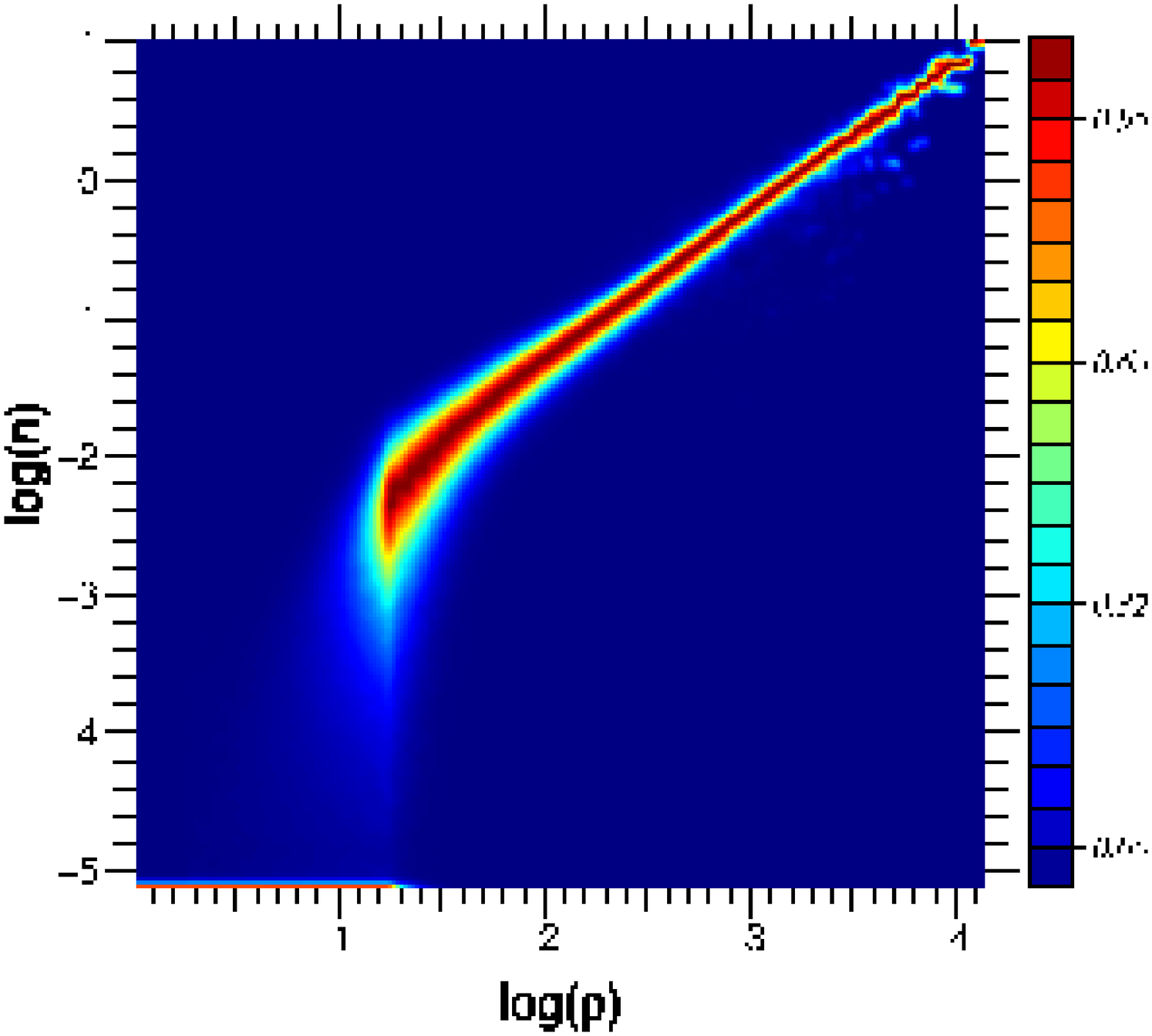}\label{fig_molusc_map_z0}}\\
  \caption{ The probability $P\sachant{n_G}{\rho_S}$ that a given galaxy
    number density $n_G$  corresponds to a dark matter mass density $\rho_S$
    for different redshifts (logarithmic scales). For low dark matter densities, no galaxy
    formation occur and the galaxy number density is thus $0$ whereas it
    appears directly proportional to the dark matter density $\rho_S$ for high
    values of $\rho_S$. In between this two regimes, a wide range of galaxy
    densities can correspond to a given dark matter density. See main text for
    an explanation of the change of behavior with redshift.\label{fig_molusc_map}}
\end{figure}

Once the sampled density fields $\rho_{S_s}\paren{\vr_i}$ and
$\rho_{G_s}\paren{\vr_i}$ are computed from $S_s$ and $G_s$, the probability
$P\sachant{n_G}{\rho_S}$  for the galaxy number density at a location $\vr$
to be $n_G\paren{\vr}$ knowing that the dark matter mass density at this same
location is $\rho_S\paren{\vr}$ can be obtained. This is achieved by applying
the following equations:
\begin{equation}
\left \lbrace
\begin{array}{l}
\displaystyle P\sachant{n_G}{\rho_S}\propto \sum_{k=1}^{N_n}\delta\paren{\rho_S\paren{\vr_k}-\rho_S}\delta\paren{n_G\paren{\vr_k}-n_G}\\
\displaystyle \int_0^\infty P\sachant{n_G}{\rho_S}\,dn_G=1
\end{array}
\right . 
\end{equation}
where the sum is computed over the $N_n$ nodes of the sampling grids (which are
identical for both distributions) and $\delta\paren{\vr}$ is the usual Dirac
function. Figure \ref{fig_molusc_map} shows the function
$P\sachant{n_G}{\rho_S}$ for three different redshifts ($z=3$, $z=1$ and
$z=0$) computed from a $512^3$ particles in a $100h^{-1}$ Mpc box dark matter
simulation and its GalICS counterpart. As expected, two different regimes
exist: when  dark matter density is low, no galaxy formation can occur whereas
for high densities, the galaxy number density is proportional to the dark
matter mass density ($n_{G_s}=b\rho_{S_s}$). Between these two regimes, a
large range of galaxy densities can correspond to a given dark matter density
depending on the galaxy formation history. At high redshifts (Figures
\ref{fig_molusc_map_z3} and \ref{fig_molusc_map_z1}), a change can be
observed for  $P\sachant{n_G}{\rho_S}$ at high $\rho_S$ values.  This can be
explained by the fact that for $z=3$, the most massive halos gravitational
collapse has not occurred yet. Hence, the halos mass function is dominated by
small size objects which cannot all be detected within the simulation due to
resolution limitations. In fact, as within the hierarchical model framework
large halos are formed by smaller halos fusions, their galaxy content is
weaker than expected at high redshifts. 

The function $P\sachant{n_G}{\rho_S}$ describes how the galaxy distribution
maps to the underlying dark matter distribution but does not contain any
information about the galaxy properties. In order to be able to attribute a
spectrum to every galaxy created from $S_l$, we compute the probability
distribution $P\sachant{F_i\paren{\lambda}}{n_G,\rho_S}$  that a given galaxy
located at point $\vr$ has a given spectrum $F_i\paren{\lambda}$, knowing that
the galaxy number density is $n_G\paren{\vr}$ and the dark matter mass density
is $\rho_S\paren{\vr}$. This distribution is obtained from $G_s$ and $S_s$,
using the $N$ synthetic spectra $F_i\paren{\lambda}$ generated by GalICS in
$G_s$:
\begin{equation}
\left \lbrace
\begin{array}{l}
\displaystyle P\sachant{F_i\paren{\lambda}}{n_G,\rho_S}\propto\\
\displaystyle \;\;\;\;\;\;\;\;\;\sum_{j=1}^{N_n}
\delta\paren{\rho_S\paren{\vr_j}-\rho_S}     \delta\paren{n_G\paren{\vr_j}-n_G}
W\paren{\vr_i-\vr_j},\\
\displaystyle \sum_{i=1}^N P\sachant{F_i\paren{\lambda}}{n_G,\rho_S}=1.
\end{array}
\right .
\label{eq_proba_spec}
\end{equation}
 In this equation, $\vr_i$ is the location of the $i^{th}$ galaxy in $G_s$,
 $\vr_j$ the coordinates of the $j^{th}$ grid node and $F_i\paren{\lambda}$
 the spectrum associated to the $i^{th}$ galaxy. This simply states that the
 probability for a given type of galaxy to exist at a place where the galaxy and
 dark matter densities are $n_G$ and $\rho_S$ is proportional to the number of
 galaxies of that type observed in $G_s$ close to grid nodes, each being
 weighted by their respective contribution to the value of $n_G$ at that node
 (hence the factor $W\paren{\vr_i-\vr_j}$). From a more practical point of
 view, a list of spectra is attributed to every pixel in Figure
 \ref{fig_molusc_map} with a weight associated to each of them.

\subsection{Generating the galaxy distribution}

The first step of the process consisted of extracting information from $S_s$
and $G_s$ through the computation of probability distributions
$P\sachant{n_G}{\rho_S}$ and $P\sachant{F_i\paren{\lambda}}{n_G,\rho_S}$. The
second step aims to build a galaxy distribution $G^{\star}_l$ out of the large scale
dark matter simulation $S_l$ using $P\sachant{n_G}{\rho_S}$ and
$P\sachant{F_i\paren{\lambda}}{n_G,\rho_S}$. To do so, we first compute the
sampled galaxy number density field $n_{G^{\star}_l\paren{\vr_i}}$
corresponding to $S_l$. Using the same technique and grid parameters as
before, the sampled mass density field $\rho_{S_l}\paren{\vr_i}$ is extracted
and, for every grid node $i$, a corresponding value of
$n_{G^{\star}_l\paren{\vr_i}}$ is randomly selected, following the probability
distribution $P\sachant{n_G}{\rho_S}$.\\

The galaxy distribution $G^{\star}_l$ can then be generated by creating a
point distribution whose sampled density field is $G^{\star}_l$. Let
\begin{equation}
\norm{W\paren{\vr}}=\sum_{i=1}^{N} W\paren{\vr-\vr_i}
\end{equation}
be the sampled norm of the kernel $W$, with $i$ the index of a grid node. The
fact that we used a {\em truncated} Gaussian kernel implies that
$\norm{W\paren{\vr}}$ is not exactly constant but rather $\sigma$-periodic. It
is nonetheless  possible to show empirically that when $L\le\sigma$, 
\begin{equation}
\displaystyle \frac{{\rm{max}}_\vr\paren{\norm{W\paren{\vr}}}-{\rm{min}}_\vr\paren{\norm{W\paren{\vr}}}}{{\rm{min}}_\vr\paren{\norm{W\paren{\vr}}}}\le
1\%,
\end{equation}
where ${\rm{min}}_\vr\paren{f(\vr)}$ and ${\rm{max}}_\vr\paren{f(\vr)}$ are
the minimal and maximal values of $f\paren{\vr}$. So as long as
$L\le\sigma$, it is possible to consider that $\norm{W\paren{\vr}}$ is a
constant and the total number of galaxies $N_{G^{\star}_l}$ in $G^{\star}_l$
is:
\begin{equation}
\displaystyle N_{G_g^\star}=\frac{1}{\norm{W\paren{\vr}}}\sum_{i=1}^{N_g} n_{G_g^\star}\paren{\vr_i},
\end{equation}
$N_g$ being the total number of grid nodes.\\

Following the hypothesis that, on the scale at which we are working (i.e of
order $1h^{-1}$ Mpc), the galaxy distribution should follow the dark matter
distribution, the galaxy distribution is generated by changing dark matter
particles from $S_l$ into galaxies or removing them following some
criteria. This way, the galaxy distribution is assured to follow the
large scale distribution of dark matter. Knowing the value of the galaxy
number density field  $G^{\star}_l$, the probability $Q_i$ for a dark matter
particle to be transformed into a galaxy can be expressed as:
\begin{equation}
Q_i\propto \frac{n_G\paren{\vr_i}}{\rho_S\paren{\vr_i}},
\end{equation}
which can be normalized using the fact that a total of  $N_{G_g^\star}$ should
be created:
\begin{equation}
Q_i=N_{G_g^\star}\paren{\frac{n_G\paren{\vr_i}}{\rho_S\paren{\vr_i}}}\paren{\sum_{j=1}{N_S}\frac{n_G\paren{\vr_i}}{\rho_S\paren{\vr_i}}}^{-1},
\end{equation} 
where $i$ corresponds to the index of one of the $N_S$ dark matter particles in
$S_l$. For every dark matter particle, located at position $\vr_i$, the local galaxy and dark matter
densities $n_G\paren{\vr_i}$ and $\rho_S\paren{\vr_i}$ are linearly interpolated from the values at
surrounding grid nodes, and every particle is changed into a galaxy or
rejected with a probability $Q_i$. The attribution of a spectrum to every
generated galaxy follows the same process: if a galaxy is created at a
location where the galaxy and dark matter density are $n_G$ and $\rho_S$, then
one of the $N_s$ spectra $F_i\paren{\lambda}$ in $G_s$ is attributed to it,
following the probability distribution
$P\sachant{F_i\paren{\lambda}}{n_G,\rho_S}$. In the unlikely case where
$Q_i>1$ (which means that the local galaxy  number density is superior to the
dark matter number density), a number of galaxies equal to the integer part of
$Q_i$ are generated and randomly located on a sphere of radius $d$ centered on
$\vr_i$, $d$ being a random number with a probability distribution
$W\paren{d}$ (i.e. identical to the kernel used for the density sampling).

\section{Verifying the Galaxy distribution}
\begin{figure*}
  \centering \subfigure[GALICS256, $40h^{-1}$ slice, $30,765$ total galaxies]{
    \includegraphics[angle=0,width=8cm]{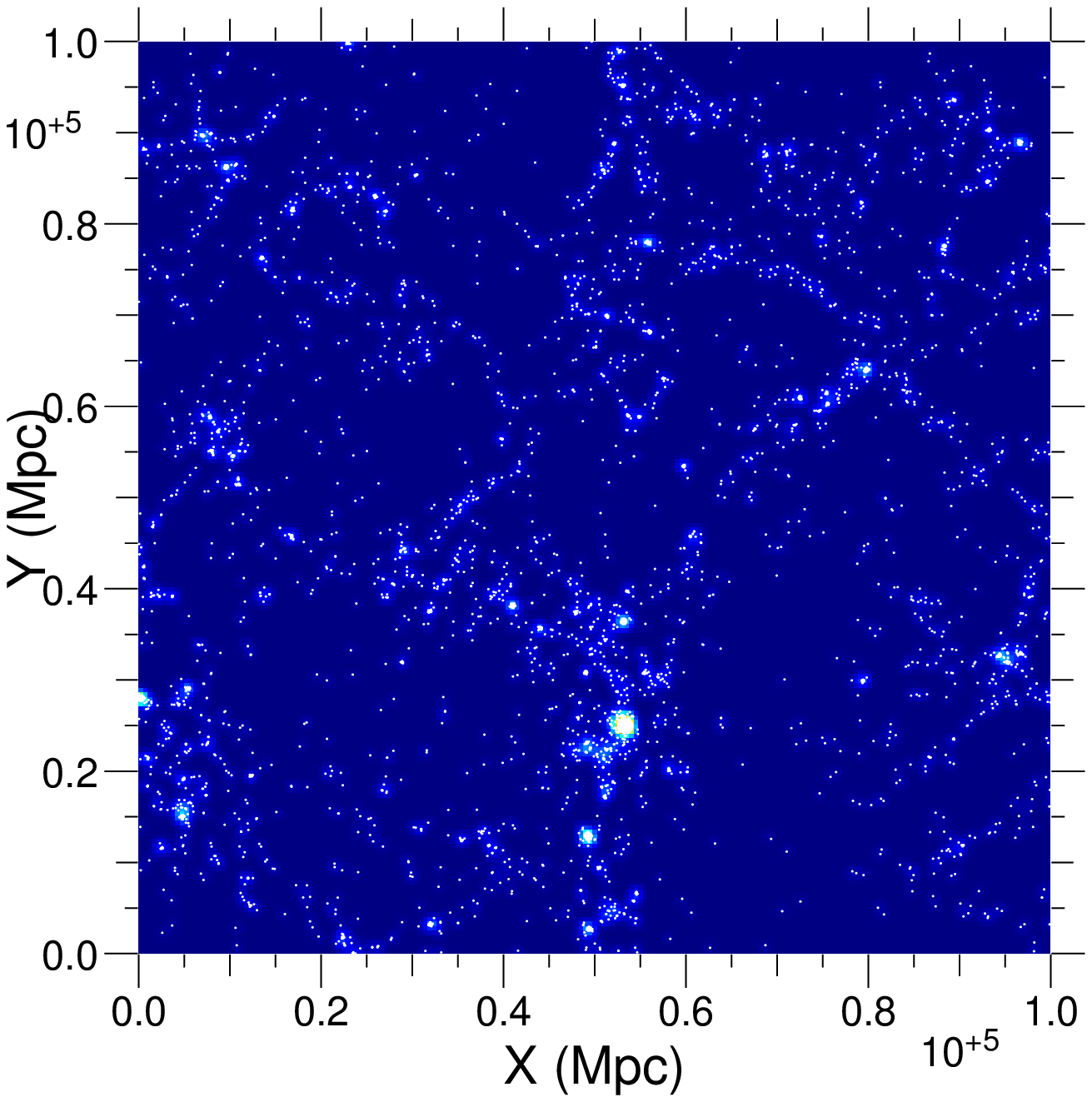}
  \label{fig:galics_snap}}
\subfigure[MOCK256, $40h^{-1}$ slice, $30,941$ total galaxies]{
  \includegraphics[angle=0,width=8cm]{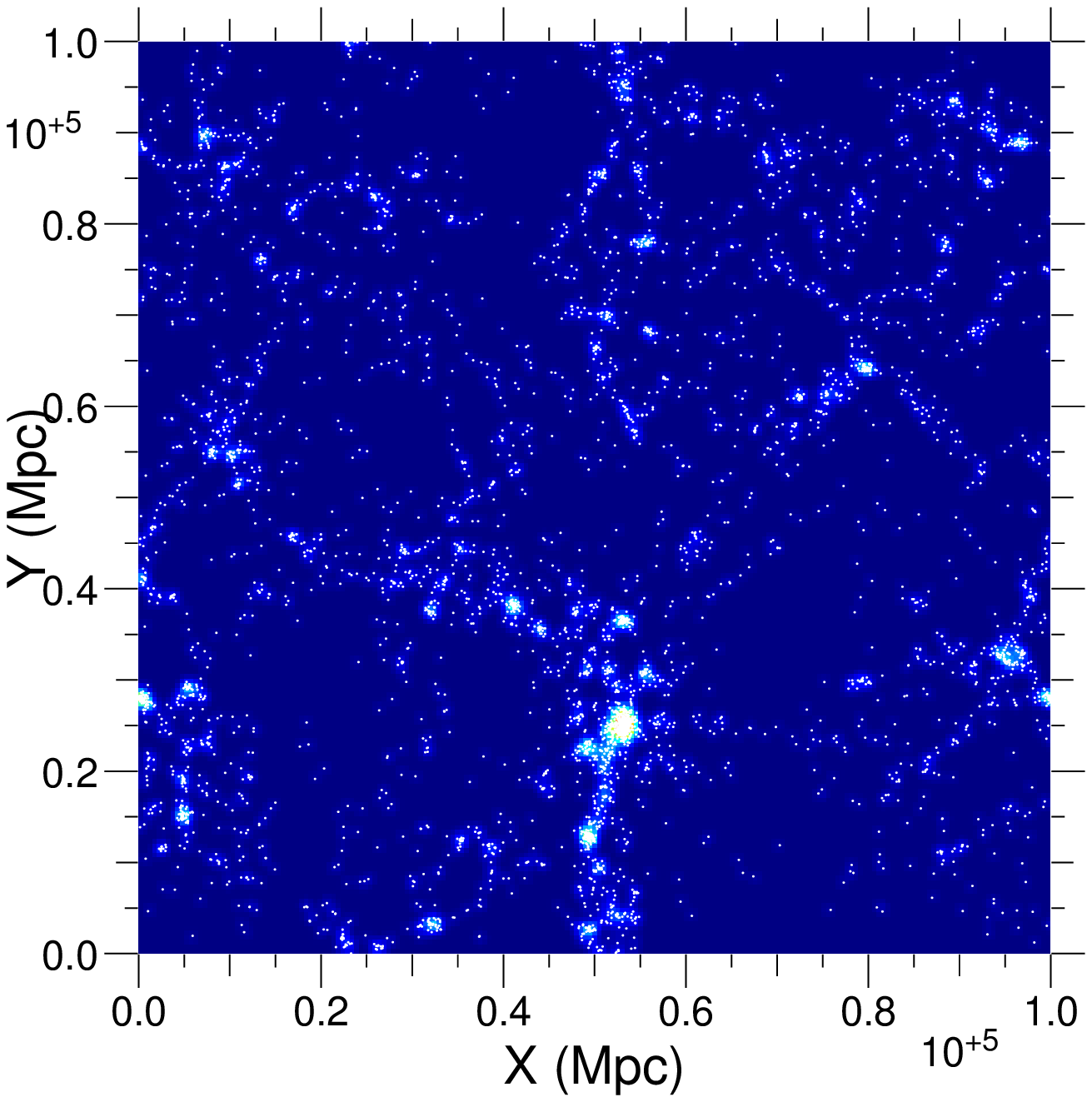}
  \label{fig:mock_snap}}
  \caption{Comparison of the two galaxy distributions obtained with GalICS and
    MoLUSC from the same dark matter distribution. The MoLUSC version was
    computed with a $1h^{-1}$ Mpc smoothing length from the dark matter
    simulation and its GalICS counterpart.\label{fig:mockcomp}}
\end{figure*}

Using the previously described process, one can generate a large scale galaxy
distribution from a large scale dark-matter-only simulation by mimicking the
properties of a smaller scale galaxy distribution generated using
sophisticated semi-analytical models such as GalICS. To check the quality of
the results, we generated a galaxy distribution $G^{\star}$ from a $100h^{-1}$
Mpc with $256^3$ particles simulation $S$ using the properties of the galaxy
distribution $G$ generated with GalICS from the {\em same} 
simulation\footnote{available from \url{http://www.galics.iap.fr}}. Figure
\ref{fig:galics_snap} shows a $40h^{-1}$ Mpc slice of $G$ and
\ref{fig:mock_snap} a $40h^{-1}$ Mpc slice of $G^{\star}$. In the two cases,
the total number of galaxies is very similar: $30,765$ with GalICS as opposed to $30,941$
for MoLUSC. Moreover, the scheme makes the distribution very similar on
scales larger than the smoothing length of $1h^{-1} Mpc$ (halos,
filaments and voids are located at the same place). Nonetheless, galaxy
clusters generated by MoLUSC clearly appear more spread out than the ones
generated with GalICS. This is due to the Gaussian smoothing that is part of
MoLUSC process. The other difference in cluster appearance is their shape but
constitute this time a strength of MoLUSC. GalICS uses a spherical collapse
approximation for halos such that galaxies are distributed completely spherically
within a halo, but this is not necessarily the case with MoLUSC which does not
require any halo identification. This results in galaxy clusters that really
follow the underlying matter distribution, thus presenting different geometries.

\begin{figure}[t]
  \centering
  \includegraphics[angle=0,width=8cm]{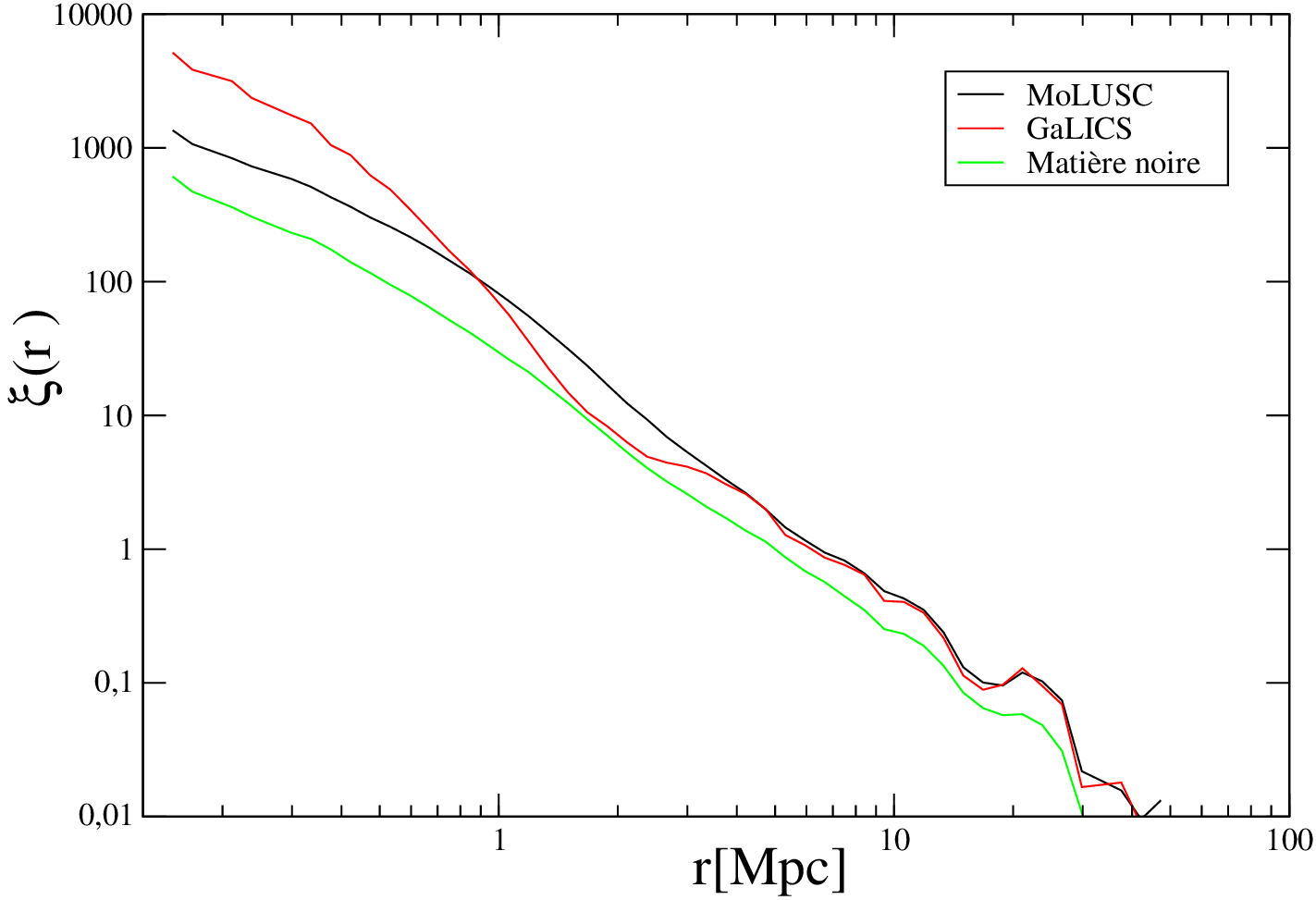}
  \caption{Two-point correlation function $\xi(r)$ 
of the dark matter distribution (green curve) and of the galaxies distributions generated by GalICS (red curve) and MoLUSC
(black curve). The smoothing length used for MoLUSC is $L=1h^{-1}$ Mpc. Part of the particles in the dark matter simulation have been randomly removed so that the number of points is the same.
}
  \label{fig:Xiplot}
\end{figure}

The examination of the two point correlation functions $\xi\paren{r}$ in
Figure \ref{fig:Xiplot} confirms the preceding remarks. MoLUSC mainly aims to
reproduce GalICS galaxy distribution. The comparison of the correlation
functions for $G$ (red curve) and the one for $G^{\star}$ (black curve)
indicates that they are similar on scales larger than $2h^{-1}$ Mpc while some
differences exist on smaller scales. It is clear that below scales of order
the smoothing length ($1h^{-1}$ Mpc here), correlations are weaker with MoLUSC
in which case they tend to follow the dark matter correlation function (green
curve). Between scales of order the smoothing length and $2h^{-1}$ Mpc, the
opposite situation appear: whereas correlations are lacking in $G$ because of
the spherical halo model used by Galics, this is not the case in $G^{\star}$.
All these observations tend to show that MoLUSC is particularly appropriate for
the production of large scale distributions of galaxy out of a large scale
dark matter-only simulation. 

\section{Real space snapshot to redshift catalog conversion}

Once the simulation has been populated with galaxies, the primary difference
remaining between a simulation and a survey like SDSS is that observing galaxies introduce biases
that are not present in a simulation. So now that we have obtained a galaxy distribution,
there is still to make a mock catalog emulating these biases. First
of all, when galaxy properties are measured with a telescope, one often only
measures their redshift and their apparent magnitude in a given filter.  All
other properties like the absolute magnitude or distance has to be computed
from these values. This of course introduce errors and limitations, the main
ones being that:
 \begin{enumerate}
 \item The spectrum of a galaxy is not constant with wavelength. So as it is
   redshifted because of the expansion of the universe, the apparent magnitude
   is not measured in the same wavelength range as the absolute magnitude.
   This can be corrected by using so called K-corrections, the value of which
   depends on galaxy morphology, redshift and filter characteristics.
 
 \item The measured redshift is used to compute a distance assuming
    it is only due to the expansion of the universe.  The peculiar
   velocities are neglected in the process and give rise to so-called
   redshift distortions and the ``finger of god" effect.
   
 \item The geometry of a survey is constrained by practical matters, giving
   rise to complex geometry as seen with SDSS on fig \ref{fig_MOCK_SDSS}.
\end{enumerate}
Some aspects of the procedure used are quite similar to the one described in
\cite{Blaizot05}. The reader can refer to this paper for further details on
some points.

To build a more realistic mock catalog, we can also take into account incompleteness
with distance, the zone of obscuration and one could imagine
to add a surface brightness selection since the absolute luminosity
is available from the simulations.
 
\subsection{Basic equations}
For a given cosmological model, the comoving distance between a galaxy at
redshift $z$ and an observer in a flat universe can be defined as follows:
\begin{equation}
D(z)=\frac{c}{H_0}\int_0^z \frac{dz}{\sqrt{\Omega_m(1+z)^3+\Omega_\phi
    f(z)}},\;\;\;\Omega_m+\Omega_\phi=1
\label{eq:distance}
\end{equation}
where $\Omega_m$ and $\Omega_\phi$ are the fraction of matter and dark energy
respectively. The function $f(z)$ depends on the cosmological model. If we
consider the $\Omega_\phi$ part to be modeled by a fluid with an equation of
state $p=w\rho$ then we obtain
\begin{equation}
f(z)=(1+z)^{3(1+w)},\;\;\; p=w\rho
\label{eq:f(z)}
\end{equation}
which gives $f(z)=1$ in the case of a LCDM model. The distance
(\ref{eq:distance}) is not directly observable, but corresponds to the
distance than one can measure directly between two galaxies in a simulation
box. It can be easily related to luminosity distance by $d_L=(1+z)D$ where
$d_L$ is the luminosity distance.

Although Equation \ref{eq:distance} is not analytically solvable in the
general case, a low redshift approximation can be used for our purpose, giving
at second order in $z$:
\begin{equation}
D(z)=\frac{c}{H_0}\left(z-\frac{1}{2}\left(1+q_0\right)z^2\right) 
\label{eq:distanceo2}
\end{equation}
with $q_0$ the deceleration parameter:
\begin{equation}
q_0=\frac{\Omega_m}{2}+\frac{\Omega_\phi}{2}\left(1+3w\right).
\label{eq:q_0}
\end{equation}
Finally the classic relationship between apparent and absolute magnitude is
given by:
\begin{equation}
M=m-5\log\left(D(z)(1+z)\right)-25+K(z)
\label{eq:mag}
\end{equation}
where $D(z)(1+z)=d_L$ is the luminosity distance, $M$ the absolute magnitude,
$m$ the apparent magnitude and $K(z)$ stands for the K-correction term.

\subsection{Mock catalog construction}
One of the major problems of studies of large-scale structure is that
it is quite difficult to compare theoretical results to
observational data in a fair way. With MoLUSC, one can generate a large scale
galaxy distribution from a dark matter only simulation. Galaxies,
being generated from an N-body simulation, are distributed in a
series of data cubes representing universe snapshots at different redshifts. A
galaxy catalog such as SDSS is different in that it represents the galaxy
distribution as viewed by an observer. Thus, making mock catalogs consists of
reproducing from the data cubes a galaxy distribution similar to what would be
observed, taking into account all the observational biases.

Initially, every snapshot contains the following information:
\begin{itemize}
\item The value of the redshift.
\item The location and velocities of all galaxies.
\item For every galaxy, a corresponding spectrum. 
\end{itemize}
By randomly placing an observer within the $z=0$ snapshot, and defining a line
of sight direction, it is possible to compute the {\em observational}
properties of the galaxies belonging to the volume of the galaxy catalog to
mimic, paying attention to the fact that the galaxies should be extracted from
the snapshot with redshift corresponding to the distance between the
observer and the galaxy. The main problem with this method is that the
original simulation can be smaller in size than the galaxy
catalog for which a mock catalog is being generated. 
Of course, one benefit of the MoLUSC technique is that
very large galaxy distributions can be easily generated, but
still it can be that every snapshot has to be duplicated in
order to obtain a distribution on a large enough volume. This technique is
called random tiling and consists of stacking simulation snapshots after
applying to each of them random rotations and translations, using the periodic
boundaries condition to ensure that the transformed snapshots have the same geometry. The random
transformations are very important in that they prevent the final distribution
from being periodic as anyway no information exists on scales larger than that
of the snapshots.  

\begin{figure*}
  \centering
  \includegraphics[angle=0,width=16cm]{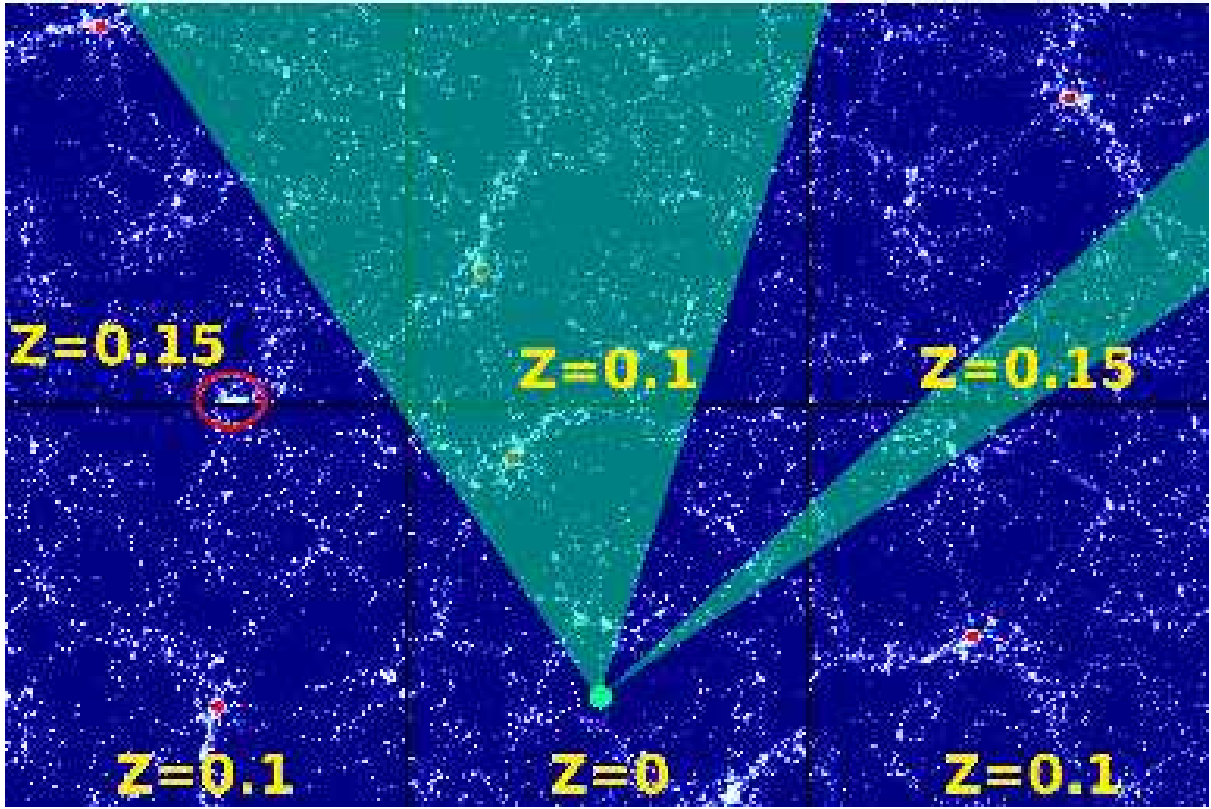}
  \caption{
Illustration of the random tiling method and of the construction of mock catalogs from
the initial catalogs. The space is made from initial catalogs with different redshifts
which have been randomly transformed (rotation, translation). The red dots show the same cluster of galaxies.
The red circle illustrates the problem of clusters lying close to a side of a box. To create a mock catalog,
only the galaxies included in the geometry of the initial catalog (here in green) are selected.
}
\label{fig:conecut}
\end{figure*}

Figure \ref{fig:conecut} illustrates the random tilling method. The main
drawback comes from the fact that a large cluster can be cut if it
is located at a place where the snapshot used to build the mock catalog
changes. This problem is solved using a friends-of-friends type structure
finder: instead of computing observational properties of the galaxies one by
one, it is done for every cluster and depending on the cluster center
position, all of its galaxies will be present or absent of the mock catalogs. The
influence of random tilling on the correlation  function can be found in
\cite{Blaizot05}, section 3.

\subsection{Final catalogs}

Galaxy catalogs often have complex geometries due to observational
constraints. It is important to reproduce them in order to be able to check
the influence of edge effects. To do so, the method we use simply consists of building a mask
from the real catalog and applying it to the mock catalog. Once the catalog is
built, the only step left involves reproducing the main observational
biases: redshift distortions and spectral redshifting. Redshift distortions are
taken into account by using the peculiar velocities of the galaxies obtained
from the simulation snapshots. To a galaxy at a given
distance we add a redshift due to hubble flow to an
additional term due to the peculiar velocity that prevents an exact
distance measurement. The amplitude of this term is simply given by:
\begin{equation}
\delta z=\sqrt\frac{1+\vv_p}{1-\vv_p}-1
\end{equation}
where $\vv_p$ is the norm of the peculiar velocity projection on the line of
sight. Finally, the galaxies' absolute and apparent magnitudes are computed
using GalICS synthetic spectra as observed through a given filter
$F\paren{\lambda}$. This way, the $i^{th}$ galaxy with redshift $z_i+\delta
z_i$ with a rest frame spectrum $P_i\paren{\lambda}$ has an observer frame
spectrum:
\begin{equation}
P^{\rm{obs}}_i\paren{\lambda} = \frac{P_i\paren{\lambda\paren{1+z_i}}}{1+z_i},
\end{equation}
and its absolute and apparent magnitudes $M_i\paren{F}$ and $m_i\paren{F}$  are: 
\begin{equation}
M_i=\int_0^\infty P_i\paren{\lambda}F\paren{\lambda}\,d\lambda
\label{eq_Mag}
\end{equation}
and
\begin{equation}
m_i=\int_0^\infty P^{\rm{obs}}_i\paren{\lambda}
F\paren{\lambda\paren{1+z_i}}\,d\lambda.\\
\label{eq_mag}
\end{equation}

\begin{figure*}
\centering
\subfigure[DR4-350]{\includegraphics[width=7cm]{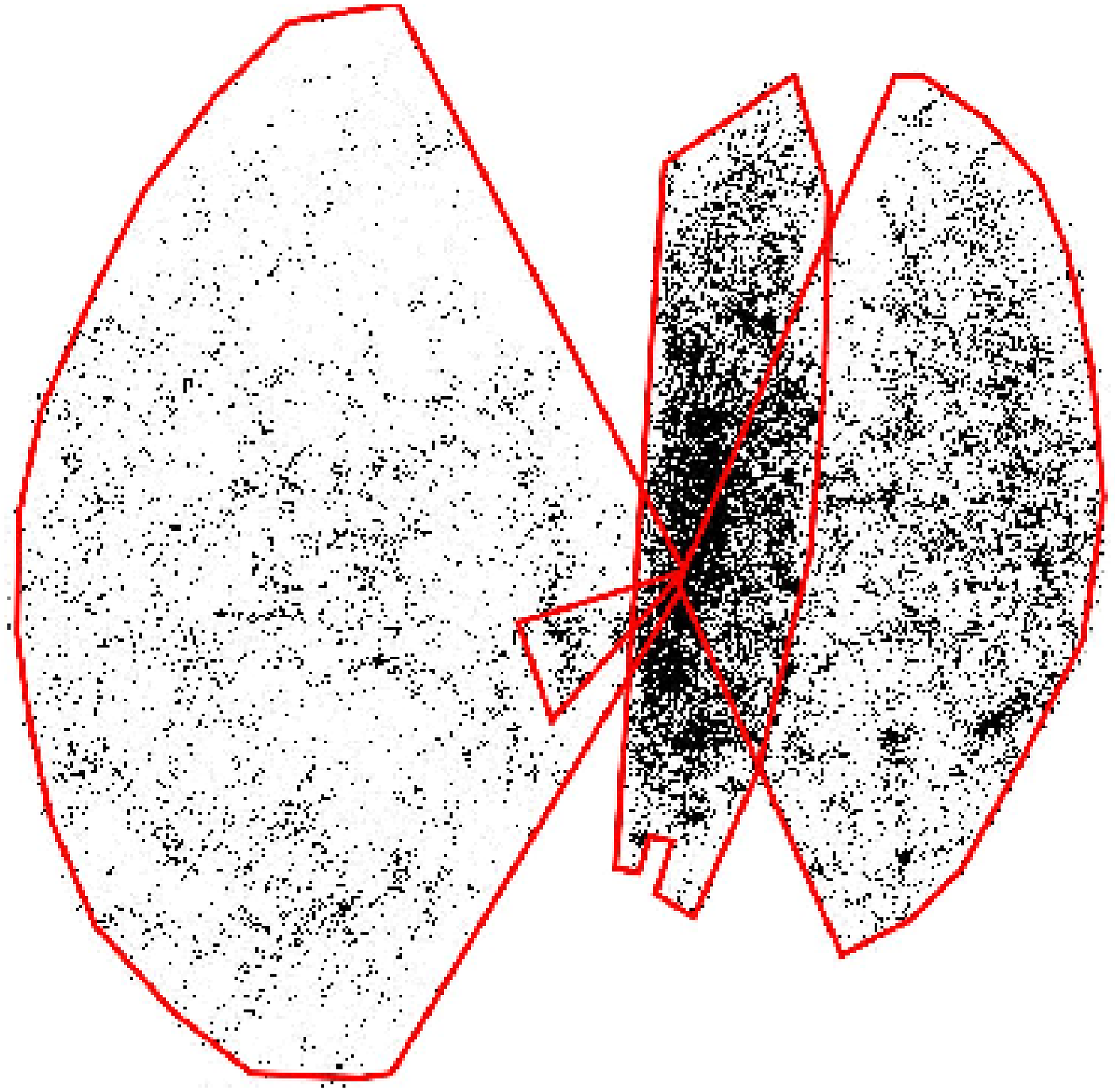}\label{fig_MOCK_SDSS_SDSS}}
\hfill\centering
\subfigure[MOCK]{\includegraphics[width=7cm]{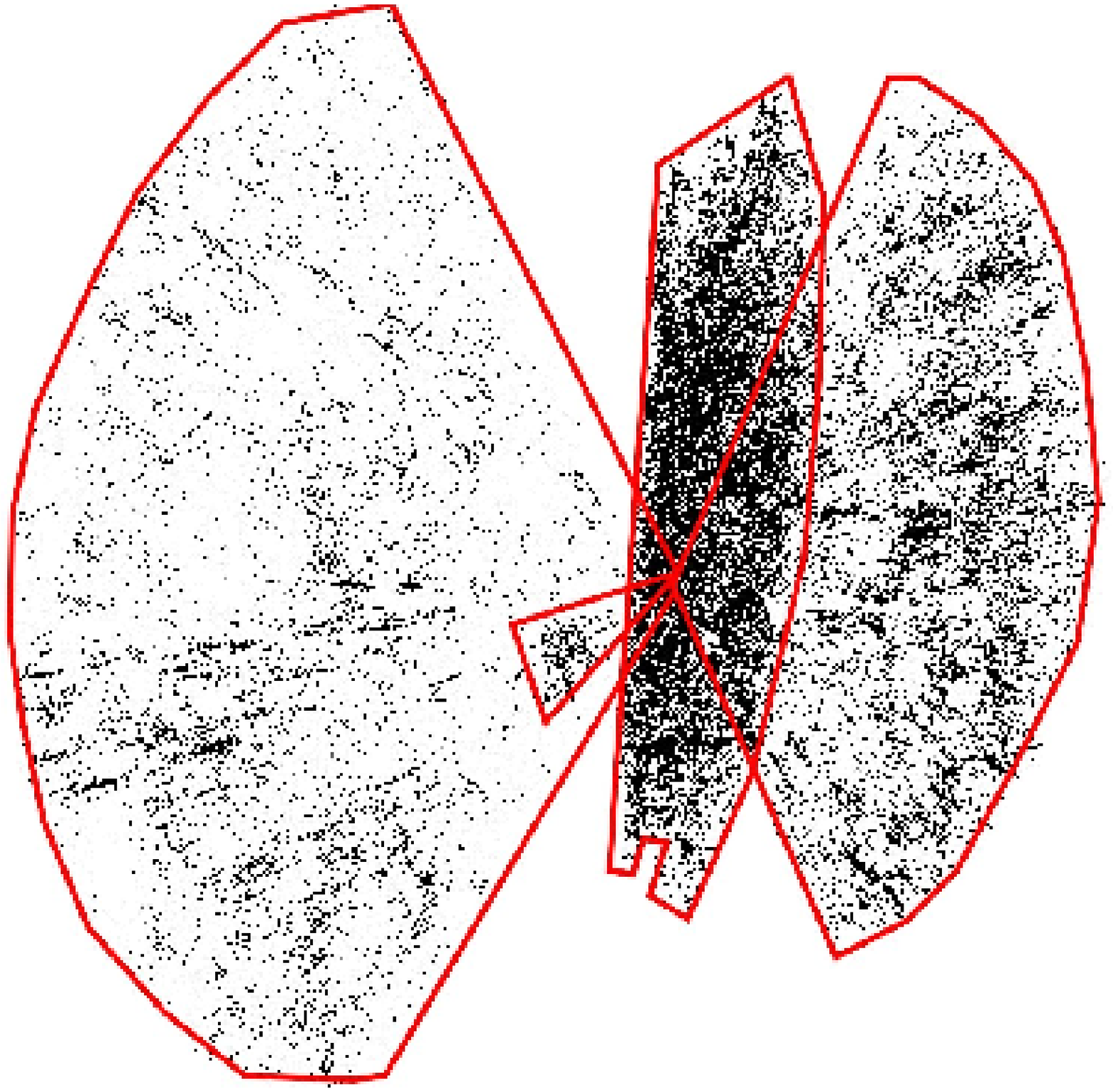}\label{fig_MOCK_SDSS_MOCK}}\\
\centering
\subfigure[MOCK-NB]{\includegraphics[width=7cm]{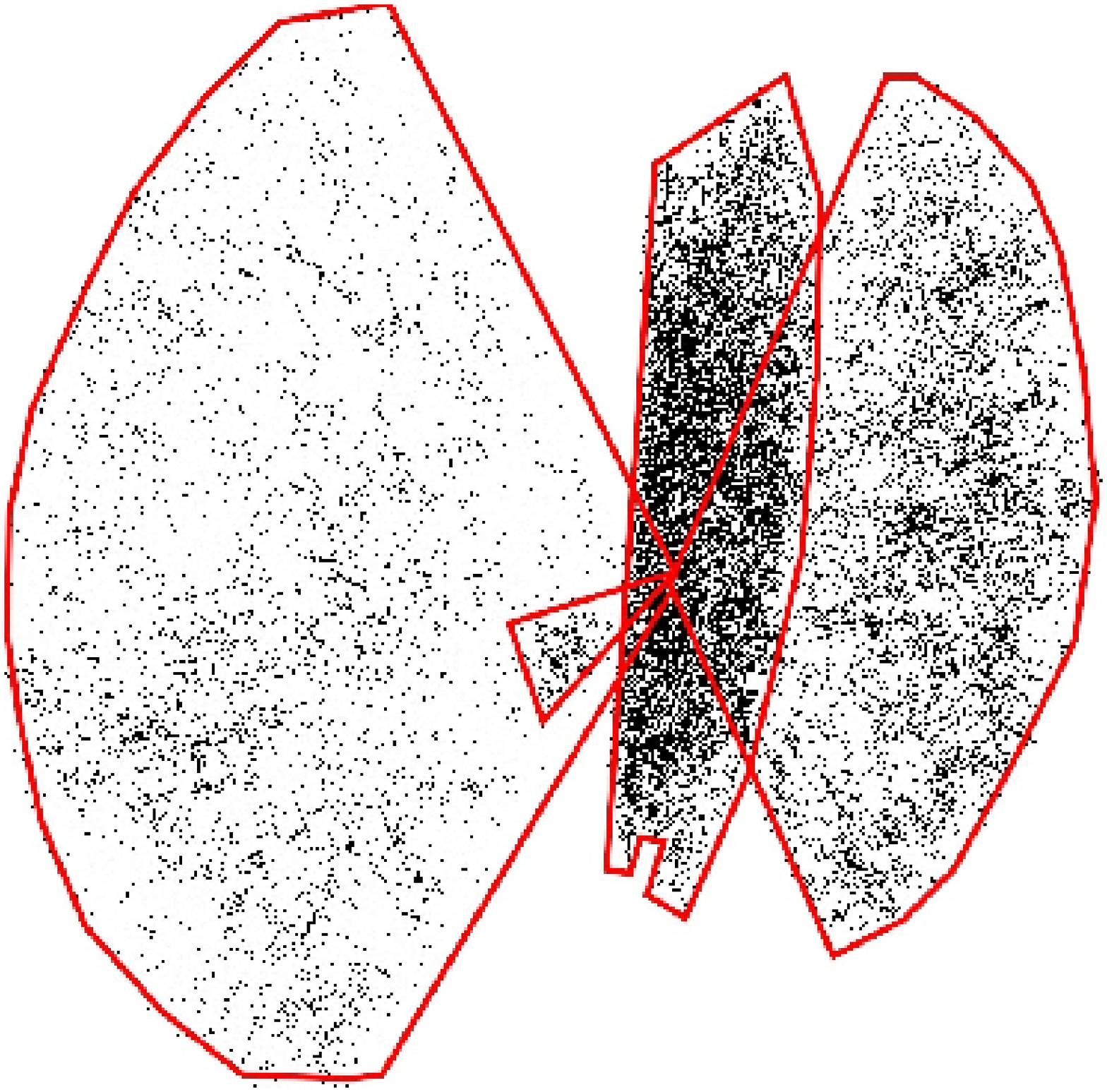}\label{fig_MOCK_SDSS_NB}}
\hfill\centering
\subfigure[MOCK-NBNF]{\includegraphics[width=7cm]{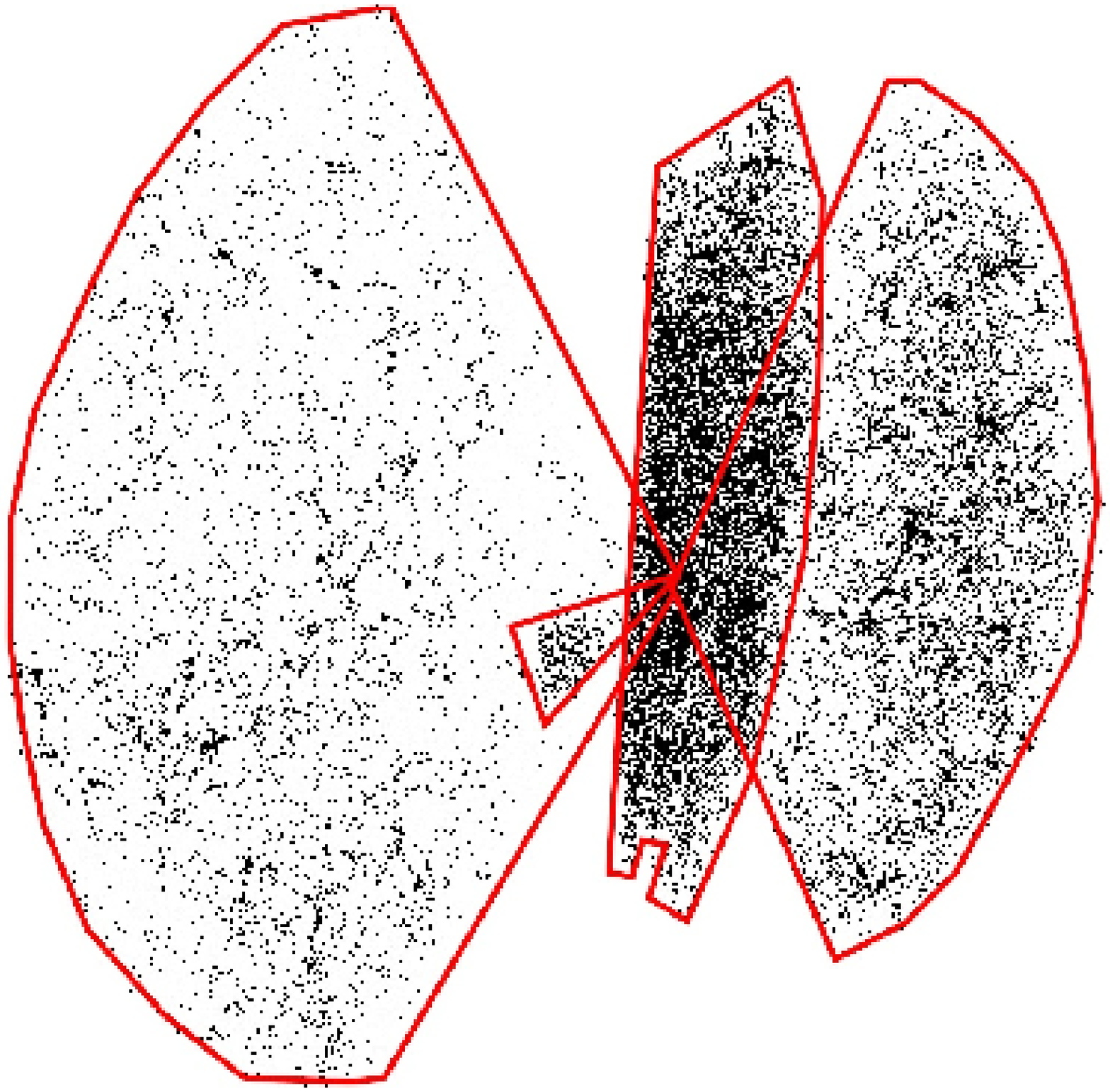}\label{fig_MOCK_SDSS_NF}}\\
\caption[Comparaisons des catalogues virtuels au SDSS]{Comparison between the
observed SDSS DR4 (a) survey (limited to 350 Mpc) and 3 different virtual SDSS
surveys (b, c, d) built from the large scale low resolution DM simulation.
The virtual catalog MOCK (b) is made taking into account all the observed
selection effect and observational artifacts as the fingers of god.  The
MOCK-NB (c) is built while neglecting the bias between galaxies and DM.  This
results in a less contrasted density field.  In MOCK-NBNF (d) both the bias
and the redshift distortions were not accounted for. As a result of this lack
of observational artifacts the galaxy clusters appear almost sphericals, not
elongated along the line of sight. \label{fig_MOCK_SDSS}}
\end{figure*}

Figure \ref{fig_MOCK_SDSS} shows the result obtained for a mock SDSS DR4
catalog using MoLUSC with different observational biases taken into account.

\section{Conclusions} 
In this paper we presented MoLUSC, a tool for extracting mock galaxy
surveys
out of large scale pure dark matter numerical simulations. Our method
consists of first
computing numerically at a given redshift how the galaxy distribution maps
to the
underlying dark matter distribution using a high resolution small scale
simulation.
We are then able to reproduce the galaxy distribution characteristics
from a snapshot of any size and resolution, while conserving
proper statistical properties.

Galaxy properties are computed as
measured by an observer. Bias effects such as fingers of god, k-corrections
on magnitudes, catalog geometry are numerically computed allowing a fair
comparison to the observed data which is crucial in order to study the
formation and evolution
of the large-scale structures in the Universe.

MoLUSC is being used to provide mock catalogs of the SDSS and of the Deep
Fields of CFHTLS,
for a variety of cosmologies. The observers are provided with mock catalogs
matching the survey
selection effects such as magnitude depth and survey geometry. The virtual
observations are provided in the corresponding
survey set of filters.

This tool provides a realistic galaxy distribution out of any large dark
matter only simulation
without using powerful computer hardware, allowing a faster
astrophysical analysis
of the present large DM simulations.

\begin{acknowledgements} 
This work was supported by the US Space Interferometer Mission Dynamics of
Nearby Galaxies (SIMDOG) Key Project,
by a european Marie Curie predoctoral grant, and by the French Horizon
Project (http://projet-horizon.fr).
Greg Bryan acknowledges support from NSF grants AST-0507161, AST-0547823, and AST-0606959.
\end{acknowledgements}


\newpage
 
\begin{thebibliography}{} 

\bibitem[Blaizot et al. 2005]{Blaizot05}
Blaizot J., Wadadekar Y., Guiderdoni B., Colombi S.T., Bertin E., Bouchet F., 
Devriendt J.E.G., Hatton S., 2005, MNRAS, 360, 159

\bibitem[Cole et al.(1998)]{Cole98} 
Cole, S., Hatton, S.,  Weinberg, D.~H., \& Frenk, C.~S.\ 1998, \mnras, 300, 945 

\bibitem[Hamana et al.(2002)]{Hamana02} 
Hamana, T., Colombi,  S.~T., Thion, A., Devriendt, J.~E.~G.~T., Mellier, Y., \& Bernardeau, F.\ 
2002, \mnras, 330, 365

\bibitem[Hatton et al. 2003]{Hatton03}
Hatton S., Devriendt J.E.G., Ninin S., Bouchet F.R., Guiderdoni B., Vibert D.,
 2003, MNRAS, 343, 75-106

\bibitem[Jing et al.(1998)]{Jing98} 
Jing, Y.~P., Mo, H.~J., \& Boerner, G.\ 1998, \apj, 494, 1 

\bibitem[Peacock \& Smith(2000)]{Peacock00} 
Peacock, J.~A., \& Smith, R.~E.\ 2000, \mnras, 318, 1144 

\bibitem[Springel et al. 2001]{Springel01}
Springel V., Yoshida N., White S. D. M., 2001, New Astronomy, 6, 51

\bibitem[Springel 2005]{Springel05}
Springel V., MNRAS 2005, 364, 1105

\bibitem[Yan et al.(2003)]{Yan03} 
Yan, R., Madgwick, D.~S.,  \& White, M.\ 2003, \apj, 598, 848 

\bibitem[Zhao et al.(2002)]{Zhao02} 
Zhao, D., Jing, Y.~P., B\"orner, G.\ 2002, \apj, 581, 876 


\end{thebibliography}
\end{document}